\documentclass[nohyper,12pt,letterpaper]{JHEP}
\usepackage{epsfig}

\newfont{\frak}{eufm10 scaled 1200}
\newcommand{\mfrak}[1]{\mbox{\frak #1}}
\newfont{\Bbb}{msbm10 scaled 1200}     
\newcommand{\mathbb}[1]{\mbox{\Bbb #1}}
\DeclareSymbolFont{AMSa}{U}{msa}{m}{n}
\DeclareSymbolFont{AMSb}{U}{msb}{m}{n}
\let\Box\relax
\DeclareMathSymbol{\Box}{\mathord}{AMSa}{"03}

\def\IZ{{\mathbb Z}}
\def\IR{{\mathbb R}}
\def\IC{{\mathbb C}}
\def\IQ{{\mathbb Q}}

\def \eqn#1#2{\begin{equation}#2\label{#1}\end{equation}}
\def \rut{2/5 transformation}
\def \abs#1{\left\vert#1\right\vert}
\def \Rut{{\mathbb G}}             
\def \lpl{L_{planck}}
\def \lst{L_{string}}
\def \ham{\mathcal H}
\def \Lag{\mathcal L}
\def \greater{>}
\def \action{\mathcal A}
\def\hacek{\accent20}                           

\title{Scherk-Schwarz SUSY Breaking in Noncommutative Field Theory}

\author{T. Banks\thanks{On Leave from Rutgers University}\\
  Department of Physics and Institute for Particle Physics\\
  University of California, Santa Cruz, CA 95064\\
E-mail: \email{banks@scipp.ucsc.edu}}

\author{W. Fischler\\
  Department of Physics \\ University of Texas,
  Austin, TX\\
E-mail: \email{fischler@physics.utexas.edu}}

\abstract{Motivated by a recent conjecture[1]that quantum
  corrections and the UV/IR connection modify the classical
relation between SUSY breaking and the cosmological constant
to the phenomenologically acceptable : $M_{SUSY} \sim M_P (\Lambda /
M_P^4 )^{1/8}$, we study SUSY breaking by boundary conditions
in noncommutative field theories.  In commutative field theory
the violations of SUSY are finite and vanish as the inverse fourth power
of the radius of the SUSY violating circle. We show that in
the noncommutative theory, as a consequence of its UV/IR connection, the
perturbative corrections to SUSY breaking are infinite.  We have not yet
performed the nonperturbative resummations to extract the
true behavior of the system.}


\received{???????? ?st, 2000} \accepted{???????? ?th, 1998}

\preprint{\hepth{0007186}\\RUNHETC-00-25\\SCIPP-24/00 \\UTTG-11-00}
\begin{document}


\section{Introduction}

In a recent paper \cite{tbfolly} one of us proposed a new
approach to the cosmological constant problem, in which the
cosmological constant is an input parameter of the
fundamental theory (related to the total number of physical
states of the universe by the Bekenstein-Hawking formula).
In addition, he proposed that SUSY is restored in the flat
space limit.  All SUSY breaking has a cosmological origin.
The problem is then to explain why SUSY breaking is so large.
Considerations based on classical SUGRA suggest the relation
$M_{SUSY} \sim \Lambda^{1/4}$, which contradicts
observational bounds.  According to \cite{tbfolly}, large
quantum gravitational renormalizations will change this
formula to $M_{SUSY} \sim M_P (\Lambda / M_P^4)^{1/8}$, which
might turn out to be compatible with experiment.

The key to this proposal is the UV/IR connection which
implies that very high energy virtual states are not
associated with short distances, and might be sensitive
to the global structure of the universe.

In this paper, we would like to present a toy model with
some of the properties described above.  We do not intend
this to be an accurate representation of nature, but merely
a cartoon which emphasizes a particular feature of the
real world.  In supersymmetric quantum field theory,
Scherk-Schwarz SUSY breaking by boundary conditions on a
circle, is
a very mild way to violate the symmetry.  In a
renormalizable field theory, it does not introduce any new
divergences.  SUSY breaking effects are finite and
calculable in terms of the renormalized parameters of the
infinite volume theory.  They vanish rapidly with the
radius of the SUSY violating circle.

Noncommutative SUSY theories in infinite volume seem to
be well behaved in perturbation theory.  In particular,
the noncommutative Wess-Zumino model in four dimensions,
which will be the focus of our study, seems to have a
sensible perturbation expansion \cite{rivelles}.
However, when we compactify on a circle (in one of the
noncommutative directions) with SS boundary conditions,
we find divergent terms in the perturbation expansion\footnote{
Note that this is not the same as studying the finite temperature
behavior of the noncommutative theory\cite{paban}, because the SS
boundary conditions are taken around one of the noncommutative directions.}.
In particular, the ground state energy, which vanishes
at infinite volume in both commutative and noncommutative
theories, and is finite on a SS circle in the commutative
theory, has an infinite contribution coming from two loop
nonplanar diagrams in the noncommutative SS compactified
theory.  Similarly, the boson-fermion energy splitting has
a divergent one loop nonplanar contribution.
The divergences originate from the UV/IR correspondence,
and the fact that, in a SS compactification, the boson, but
not the fermion, has a zero momentum mode around the circle.

In previous studies of divergences in noncommutative field
theory \cite{msvretal}, it was argued that the divergences
encountered in nonplanar graphs were an artifact of
perturbation theory and could be eliminated by an adroit
resummation.  We have attempted to apply that wisdom to the
present model, but so far without success.  Simple Dyson
resummation of propagator graphs does not eliminate the
vacuum energy divergences.  As a consequence, we do not
yet know the correct nonperturbative behavior of our model.
However, we believe that our perturbative computations
establish the fact that UV/IR conspiracies can dramatically
enhance the effect of SUSY breaking in the global
structure of spacetime.

Our calculations are presented in the next section.  In the
conclusions we propose a large $N$ version of our model
which might allow us to extract the nonperturbative
corrections.  We point out however that the attempt to
extract the nonperturbative behavior of the model may
run afoul of the well known triviality of the WZ
model.  Although we have verified that very similar
divergences occur in SS compactified ${\cal N} = 4$ Super
Yang-Mills theory, we have no scheme at hand for
estimating its nonperturbative behavior

Finally, we wish to emphasize that the model of the present paper
is just a toy which reproduces a single qualitative property of the
physics conjectured in \cite{tbfolly} - namely the enhancement of
soft SUSY breaking due to the UV/IR connection.  We do not expect to calculate
the critical exponent relating the cosmological constant to the size of
SUSY breaking, in this paper.  Furthermore, it is
extremely important to realize that the vacuum energy we compute in
noncommutative field theories has no connection to the cosmological
constant.  It is merely an order parameter for SUSY breaking in a theory
decoupled from gravity.

\section{\bf Calculations}

\subsection{Calculation of the vacuum energy}

The component Lagrangian of our model is
\eqn{lag}{{\cal L} = i \partial_{\mu} \psi^* \bar{\sigma}^{\mu}
\psi + \partial_{\mu}\phi^* \partial_{\mu} \phi - 1/2 \psi
\psi - 1/2\psi^* \psi^* - g \psi\star\psi\star\phi -
g \psi^* \star \psi^* \star \phi^* - \vert m\phi + g
\phi\star\phi \vert^2 .}
The one loop vacuum graphs are purely planar and give the
same results in commutative and noncommutative WZ models.
They are finite and of order $1/R^4$.  At two loops
there are both planar and nonplanar graphs.  They have the
same combinatorics, and differ only in the presence or absence
of a Moyal phase.  We will write only the expressions for
the nonplanar diagrams.  There are three of them at this
order, given by Fig. 1 (a,b,c).

\vskip1cm
\begin{center}
\epsfig{file= 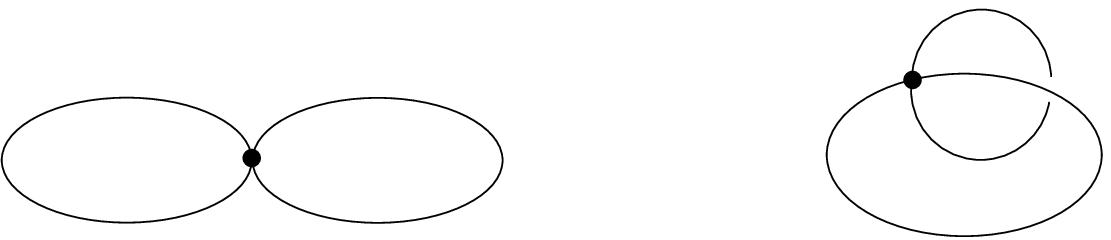} 
\end{center}

\centerline{ {\bf Fig.1a:} Planar and nonplanar two loop contributions to the
vacuum energy.}

\vskip1cm

\vskip1cm
\begin{center}
\epsfig{file= 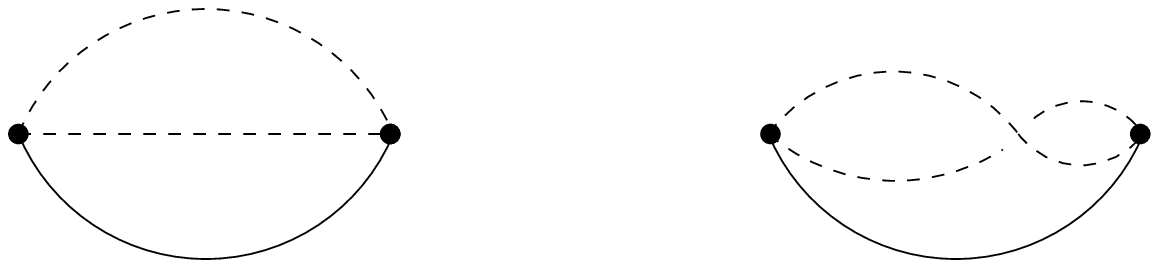} 
\end{center}

\centerline{ {\bf Fig.1b:} Planar and nonplanar two loop contributions to the
vacuum energy.}

\vskip1cm
\vskip1cm
\begin{center}
\epsfig{file= 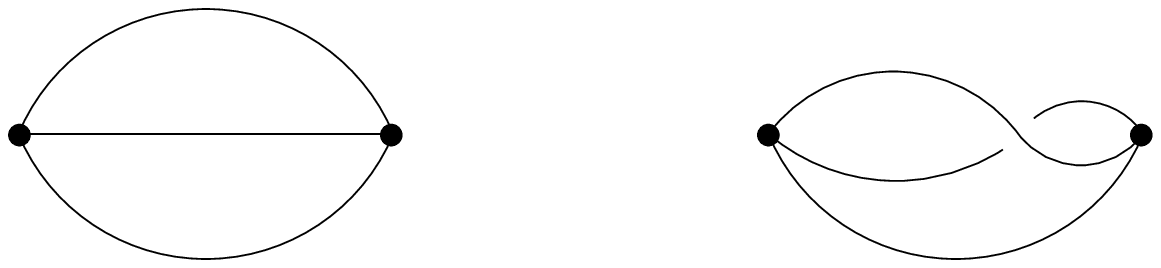} 
\end{center}

\centerline{ {\bf Fig.1c:} Planar and nonplanar two loop contributions to the
vacuum energy.}

\vskip1cm

Each has a factor of
${1\over R^2}\int {d^3 p \over (2\pi)^3}
\int {d^3 q \over (2\pi)^3}$.  The integrands are (bold face letters
denote vectors of the three continuous momenta. Scalar products are
Minkowskian, with positive time signature.  An $i\epsilon$ prescription
is left implicit.):

\eqn{npvaca}{2i g^2 \sum_{(n,l)} {e^{i{\theta\pi\over R}(
2nq_2 - 2lp_2 )}\over ({- 4n^2\over R^2} +
 {\bf p}^2 - m^2)({- 4l^2\over R^2} + {\bf q}^2 - m^2)}.}

\eqn{npvacb}{2i m^2 g^2 \sum_{(n,l)} {{e^{i{\theta\pi\over R}(
2nq_2 - 2lp_2 )}}\over {({- 4n^2\over R^2} + {\bf p}^2 - m^2)({- 4l^2\over
 R^2} + {\bf q}^2 - m^2)({- 4(l+n)^2\over R^2} + { ({\bf p} + {\bf q})}^2 - m^2)}}.}

\eqn{npvacc}{4i g^2 \sum_{(n,l)}{ {({\bf pq} - {(2n+1)(2l+1)\over R^2})
 {e^{i{\theta\pi\over R}(
(2n+1)q_2 - (2l+1)p_2 )}}}\over {({- (2n+1)^2\over R^2} + {\bf p}^2 -
m^2)({- (2l+1)^2\over
 R^2} + {\bf q}^2 - m^2)({- 4(l+n+1)^2\over R^2} +
({\bf p} +{\bf q})^2 - m^2)}}.}

Allowing shifts of momenta, and using momentum conservation
and the antisymmetry of the Moyal phase, we can rewrite the
sum of these integrands as

\begin{eqnarray}
{2i g^2 \over R^2}\sum_{n,l} &[& {e^{i{\theta\pi\over R}(2nq_2 -
2lp_2)}\over ({- 4n^2\pi^2 \over R^2} + {\bf p}^2 - m^2)({-
4l^2\pi^2 \over R^2} + {\bf q}^2 - m^2)}  \nonumber
 \\
 &-& 2 {e^{i{\theta\pi\over R}(2nq_2 - (2l+1)p_2)} \over
({- 4n^2\pi^2 \over R^2} + {\bf p}^2 - m^2)({- (2l+1)^2\pi^2 \over R^2}
+ {\bf q}^2 - m^2)}
\nonumber \\
&+& {e^{i{\theta\pi\over R}((2n+1)q_2 - (2l+1)p_2)} \over
({- (2n+1)^2\pi^2 \over R^2} + {\bf p}^2 - m^2)({- (2l+1)^2\pi^2 \over R^2} +
{\bf q}^2 - m^2)}]
\end{eqnarray}

This is all to be integrated over three momenta with the
usual phase space.   Note that if it were not for the mismatch
between the quantization rules for fermionic and bosonic
momenta, the integrand would vanish.
\vfill\eject
Doing the integral over momenta, we obtain the two loop
nonplanar contribution to the vacuum energy:

\begin{eqnarray}
E/V = {2  g^2 \over 16\pi^2\theta^2}\sum_{(n,l)} &[ &{e^{-\vert
{\theta\over R}\vert(\vert 2n \sqrt{4l^2 /R^2 + m^2}\vert + \vert
2l\sqrt{4n^2 /R^2 + m^2}\vert )} \over 4\vert n l \vert}\nonumber
\\ & - &  2 {e^{-\vert {\theta\over R}\vert(\vert 2n
\sqrt{(2l+1)^2 /R^2 + m^2}\vert + \vert 2l\sqrt{(2n+1)^2 /R^2 +
m^2}\vert)} \over \vert 2n (2l+1)\vert}\nonumber \\ & + &
{e^{-\vert {\theta\over R}\vert(\vert (2n+1) \sqrt{(2l+1)^2 /R^2 +
m^2}\vert + \vert (2l+1)\sqrt{(2n+1)^2 /R^2 + m^2}\vert)} \over
\vert (2n+1) (2l+1)\vert}].
\end{eqnarray}

The terms in this sum where either or both of $n$ and $l$
vanish, are quite divergent, and the divergence does not cancel.  The
sum over large values of the discrete momentum is quite
convergent, and Bose-Fermi cancellation is also restored.

\subsection{Calculation for the masses}

Another indicator of SUSY breaking is the spectrum of
elementary excitations.  There are three nonplanar one loop
graphs which contribute to the boson self energy (Fig. 2 a,b,c)
and one fermion self energy graph (Fig. 3)in the same order.

\vskip1cm
\begin{center}
\epsfig{file= 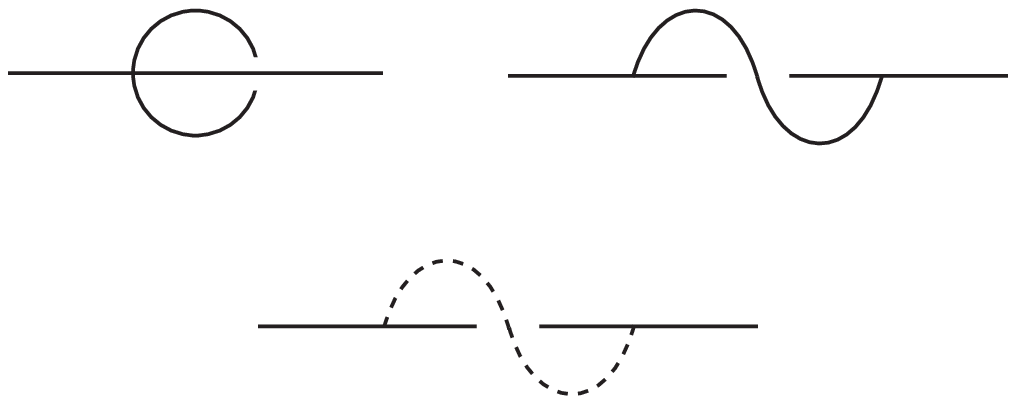} 
\end{center}

\centerline{ {\bf Fig.2:} Nonplanar one loop
graphs which contribute to the boson self energy.}

\vskip1cm

\vskip1cm
\begin{center}
\epsfig{file= 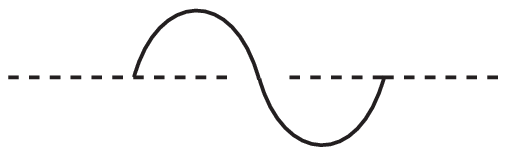} 
\end{center}

\centerline{ {\bf Fig.3:} Nonplanar one loop
graphs which contribute to the fermion self energy.}

\vskip1cm

We begin with the boson self energy.  The integrands of
the three diagrams are

\begin{eqnarray}
I_{B1} &=& \hskip1.5in {4e^{2i{\pi\theta\over R}(nk_2 -
lp_2)}\over {- 4{\pi}^2l^2\over R^2} + {\bf k}^2 - m^2}
 \nonumber \\
I_{B2} &=& \hskip1.5in {- 4 e^{{i\pi\theta\over R}(2nk_2 -
(2l+1)p_2)}\over - {\pi^2(2l+1)^2\over R^2} + {\bf k}^2 -
m^2}
\nonumber \\ & + & {e^{{i\pi\theta \over
R}(2nk_2 - (2l+1)p_2)}({- 8n^2 \pi^2\over R^2} +{2\bf p}^2
-4m^2)\over ({- {\pi}^2(2l+2n+1)^2 \over R^2} + ({\bf p}+{\bf k})^2 -
m^2)({- {\pi}^2 (2l+1)^2\over R^2} + {\bf k}^2 - m^2)}
\nonumber \\
I_{B3} &=& \hskip.5in {4m^2 e^{{2i\pi\theta\over R}(nk_2 -
lp_2)}\over ({- 4{\pi}^2l^2\over R^2} + {\bf k}^2 - m^2)({- 4\pi^2
(n+l)^2 \over R^2} + ({\bf p} + {\bf k})^2 - m^2)}
\end{eqnarray}
The dispersion relation for the bosons has singularities, in stark
contrast to the dispersion relation for fermions, as will be shown in what
follows.

 The singular behaviour of the boson dispersion relation
occurs for
 special momenta along the noncommuting directions $x_2$
and $x_3$. These singularities appear when integrating
over the large Fourier components of the various internal lines contributing to
the two-point function of bosons.The external momenta
at which these singularity occur are: $p_3 = n/R =0$ and $p_2 =
n'R/\theta$ , where $n$ and $n'$ are integers.

This can be seen when one adds the three diagrams (Fig 2 a,b,c):
                                                                             
\eqn{gamma} {\Gamma_{B}(p) = {{1\over R}{\sum_{l}\int {d^3 p \over
(2\pi)^3}(I_{B1} +I_{B2} +I_{B3})}}},
where $\Gamma_{B}(p)$ is the nonplanar contribution to the 1PI two-point
function.

Indeed it is the large l part of the sum and large $\bf k$ region
of integration in the integrals that appear in the previous equation,
that are responsible for the singularities. For these value of the
external momenta, the Moyal phase factor disappears and therefore the
large momenta contributions remain unsuppressed, as in the case of planar
diagrams.

One can single out two divergent contributions to $\Gamma(p)$.
The first one is:

\eqn{p^2} {{g^2\over 2\pi}({\bf p}^2 - { 4n^2{\pi}^2 \over R^2})log(1 -
2e^{-4{\pi}^2\vert\theta n\vert \over R^2} cos{2\pi\theta p_2\over R} +
e^{-8{\pi}^2\vert\theta n\vert \over R^2})\vert_{n=0,p_2=n'R/\theta}}

This expression is logarithmically divergent at $n = 0$ and
$p_2=n'R/\theta$.

When evaluated at $\theta = 0$ for arbitrary external momenta,
expression (2.11) becomes the usual nonplanar
 contribution to the wave function renormalization of
the scalars, in the commutative Wess-Zumino model.

The terms proportional to $ m^2$ coming from of (2.7) and (2.9) to (2.10)
are not singular at $n=0$.This is related to the fact that in the
commutative limit,
$\theta = 0$, the superpotential is protected by a non-renormalization
theorem.

The remaining divergent contribution to the two-point function is:

\eqn{thetaminusone} {{g^2\over \pi \vert\theta n\vert }\sum_l(e^{{-2\vert
\theta n\vert\over R}({m^2 + {4l^2\over R^2}})^{1/2} -{2i\pi \theta lp_2
\over R} } - e^{{-2\vert
\theta n\vert\over R}({m^2 + {(2l+1)^2\over R^2}})^{1/2} -{i\pi \theta
(2l+1)p_2
\over R} })\vert_{n=0,p_2=(2n'+1){R\over \theta}}}
where $n'$ is an integer.

This has a power law divergence in $n$ at $n=0$ and $p_2=(2n'+1){R\over
\theta}$. This somewhat peculiar formula resembles the formula for string
windings around a circle of radius R with a string tension $\theta$. We
note however, that it is $x_3$ that is compact and not $x_2$, so the
notion of winding around the $x_2$ direction does not seem to make sense.

A more reasonable explanation of the quantized values of $p_2$ is that
$x_3/R$ is an angle. Thus, the conjugate dimension $x_2$ is
in some sense quantized in units of $\theta\over R$.
This in turn implies that $p_2$, the
momentum conjugate to $x_2$, is an angular variable so that
expressions involving $p_2$ must be periodic with period $R\over \theta$.

These infrared divergences cannot be treated by simply resumming a
geometric series in the 1PI two-point function. Indeed this procedure does
not resolve, in the case of non-commuting compact dimensions, the infrared
singularities that occur in the vacuum energy, although it does displace
the apparent $n=0$ singularities of the zeroth order propagator to
infinity.  The latter observation was also made in \cite{gomis}.
However, this is not sufficient to make the vacuum energy finite.  Note
that in a SUSY violating ({\it e.g.} purely bosonic) noncommutative
field theory in infinite spacetime, the
measure of integration over momenta gives a finite vacuum energy after
Dyson resummation of the lowest order nonplanar self energy.
With SS compactification, the more serious IR divergences of the bosonic
zero mode remain even after Dyson resummation, despite the fact that
the geometric series in the 1PI two-point function generates a well
behaved dispersion relation.  One way to render these diagrams finite
would be to add a phase, $\eta$, to the boundary conditions for both bosons and
fermions.  This may be a useful intermediate regulator, but we really
want to understand the $\eta = 1$ limit, where the IR behavior is singular.
In the Conclusions we will suggest a
systematic, large N approach to the IR behavior of this model, which may
give a finite answer for the vacuum energy.

The fermion self-energy, $\Gamma_{F}(p), $ is given by the diagram
depicted in Fig 3.

\eqn{GammaF}{ \Gamma_{F}(p) = -{4g^2\over R}\sum_{l}\int {d^3 p \over
(2\pi)^3}{{-(2n+4l+1)\pi\tau_3\over R} + ({\bf p} +
{\bf k}) {\bf \tau}\over
[-({2n+2l+1\over R})^2 \pi^2 +({\bf p}+{\bf k})^2 - m^2][-{4l^2\pi^2\over
R^2} +{\bf k}^2-m^2]}e^{i{\theta\pi\over R} [(2n+1)k_2 - 2lp_2]}}
where the $\tau_i$ are Pauli matrices and $\tau_0$ is the identity.

The potential singularities for the fermionic two-point function,
as in the case of the scalars, originate in the region of large l for the sum
and large k for the integral.

The contribution to
$\Gamma_{F}(p)$ from this region is:

\eqn{GammaFasymp} {\Gamma_{F}(p) \sim
- {4g^2\over2\pi}({-(2n+1)\pi\tau_3\over R}+{\bf p} {\bf\tau})\log(1-2e^{-{2\pi^2\vert (2n+1)\theta \vert\over R^2}}{
cos{2\pi\theta p_2\over R} + e^{-4\pi^2\vert (2n+1)\theta \vert\over
R^2})}}

In contrast to the bosonic case,this contribution is non-singular at $n=0$.
Notice that $\Gamma_{F}(p)$ is singular in the large R limit as it should
be, since this gives the nonplanar contribution, in the
commutative limit, to the wavefunction renormalization of the fermions.

The extreme sensitivity to SUSY violating boundary conditions which we
have encountered here is unlike anything in local quantum field theory.
Here is a
set of words which captures the spirit of what is going on:  Individual graphs
in a local supersymmetric theory have UV divergences which cancel
between bosons and fermions.  In the noncommutative theory, these can be
converted into IR divergences by the UV/IR connection.  With SS boundary
conditions, bosonic lines have discrete zero modes, while fermions do
not, so the IR divergences do not cancel.  Thus, quantum corrections to
SUSY breaking are enhanced because the high energy states have large
extent in the SS direction, and are thus sensitive to SUSY
breaking.   In local field theory, high energy states have very short
wavelengths in the SS direction and are insensitive to the boundary
conditions.

\section{\bf Conclusions}

We believe the calculations that we have already done put the
conjectures of \cite{tbfolly} on a much firmer footing.  We have
exhibited a system where SUSY breaking via the large scale structure of
spacetime seems to be much larger than local field theory would have led
us to expect.  Nonetheless, one would be more comfortable if we had
obtained finite answers.  Usually (meaning of course in local field
theory), IR divergences represent merely the
breakdown of perturbation theory and appropriate nonperturbative
approximations give finite answers which do not have an asymptotic
expansion in the couplings.

We have seen that the simple Dyson resummations of
\cite{msvretal}\cite{gomis} do not give us a finite vacuum energy.
We would like to propose instead a large N resummation of the model.
The simplest model of this type has a singlet chiral superfield $\Sigma$
and and $O(N)$ vector $\Phi^i$, with superpotential
\eqn{W}{W = \mu^2 \Sigma^2 + m^2 (\Phi^i )^2 + g \Sigma (\Phi^i )^2.}
We can make it noncommutative and impose SS boundary conditions
in a manner precisely analogous to what we have done for the single
field model.   The large N approximation leads to a very complicated set
of self consistent integral equations, which we have not yet mastered.
The gap functions must be chosen to depend on the momenta in the
noncommutative directions.   We hope to report on approximate
solutions of these equations in the future.

We note however the possibility that this line of research may run into
difficulty.  The commutative WZ model does not really have an
interacting continuum limit.  Since the planar diagrams in the
commutative and noncommutative theories are identical, it is hard to
imagine that the noncommutative theory is any better behaved in the UV.
On the other hand, in the noncommutative theory UV divergences get
transformed into IR divergences.  The standard argument that IR
divergences are only an artifact of perturbation theory seems somewhat
shaky in the context of a theory whose nonperturbative existence is
doubtful.  Still, some hope can be gleaned from the remark that the
power law IR divergences we have encountered are not obviously connected
to the coupling constant renormalization which renders the model
trivial.  Only a full investigation of the large N model can produce a
conclusive answer to this puzzle.

We do believe that it is likely that many SUSY violating noncommutative
field theories simply do not exist.  That is, their IR behavior is not
independent of high energy physics.   Thus, were it not for the high
degree of supersymmetry of the model they studied, we would be
suspicious of the decoupling arguments of \cite{cdssw}.
One might regard the absence of a continuum limit for SUSY violating
noncommutative field theories as a baby version of the result conjectured in
\cite{tbfolly}, namely that there are no asymptotically flat SUSY
violating vacua of M theory.

If the large N WZ model is ill defined, we might have to turn to $N=4$
SYM theories to study the phenomenon discovered in this paper.  Here the
problem is that there is no known soluble nonperturbative approximation
which respects gauge invariance.  Perhaps the AdS/CFT correspondence or NCOS
theory\cite{ncos} can shed some light on this subject.

\section{\bf Acknowledgments}
The research of W.F. was supported in part by the Robert A. Welch Foundation
and NSF Grant PHY-9511632. The work of T.B. was supported in part by the
Department of Energy under grant DE-FG02-96ER40559.
W.F. would like to thank the Rutgers NHETC for their kind hospitality
during which time most of this work was done. In particular, he would like to
acknowledge O.Aharony, M.Douglas, G. Moore, A.Rajaraman and M.Rozali for
very stimulating discussions.  This work began while both authors
were visiting the Stanford Institute for Theoretical Physics, and we
would like to thank the members of the Institute for their hospitality.
In particular, we would like to thank L.Susskind for suggesting that
NCFT might provide a laboratory for studying the relation between SUSY
breaking and the UV/IR connection.  Discussions with E.Martinec during
this period are also gratefully acknowledged.

%


\end{document}